\documentstyle[12pt,epsfig]{article}

\newcommand{\be}{\begin{equation}}
\newcommand{\ee}{\end{equation}}
\begin{document}  
\topmargin 0pt
\oddsidemargin=-0.4truecm
\evensidemargin=-0.4truecm
\renewcommand{\thefootnote}{\fnsymbol{footnote}}
\newpage
\setcounter{page}{0}
\begin{titlepage}   
\vspace*{-2.0cm}  
\begin{flushright}
hep-ph/0608049v3\\
\end{flushright}
\vspace*{0.1cm}
\begin{center}
{\Large \bf Two Gallium data sets, spin flavour precession and KamLAND
} \\ 
\vspace{0.6cm}

\vspace{0.4cm}

{\large 
Bhag C. Chauhan\footnote{On leave from Govt. Degree College, Karsog (H P) 
India 171304. E-mail: chauhan@cftp.ist.utl.pt},
Jo\~{a}o Pulido\footnote{E-mail: pulido@cftp.ist.utl.pt}\\
\vspace{0.15cm}
{  {\small \sl CENTRO DE F\'{I}SICA TE\'{O}RICA DAS PART\'{I}CULAS (CFTP) \\
 Departamento de F\'\i sica, Instituto Superior T\'ecnico \\
Av. Rovisco Pais, P-1049-001 Lisboa, Portugal}\\
}}
\vspace{0.25cm}
and \\
\vspace{0.25cm}
{\large Marco Picariello\footnote{E-mail: Marco.Picariello@le.infn.it}} \\
\vspace{0.15cm}
{\small \sl  I.N.F.N. - Lecce, and 
Dipartimento di Fisica, Universit\`a di Lecce\\
Via Arnesano, ex Collegio Fiorini, I-73100 Lecce, Italia}

\end{center}
\vglue 0.6truecm
\begin{abstract}
We reexamine the possibility of a time modulation of the low energy solar neutrino 
flux which is suggested by the average decrease of the Ga data in line with our
previous arguments. We perform two separate fits to the solar neutrino data, one 
corresponding to 'high' and the other to 'low' Ga data, associated with low
and high solar activity respectively. We therefore consider an alternative to 
the conventional solar+KamLAND fitting, which allows one to explore the much wider 
range of the $\theta_{12}$ angle permitted by the KamLAND fitting alone. We 
find a solution with parameters $\Delta m^2_{21}=8.2\times 10^{-5} eV^2, 
tan^{2}\theta=0.31$ in which the 'high' and the 'low' Ga rates lie far apart 
and are close to their central values and is of comparable quality to the global 
best fit, where these rates lie much closer to each other. This is an indication 
that the best fit in which all solar and KamLAND data are used is not a good 
measure of the separation of the two Ga data sets, as the information from the low 
energy neutrino modulation is dissimulated in the wealth of data. Furthermore 
for the parameter set proposed one obtains an equally good fit to the KamLAND 
energy spectrum and an even better fit than the 'conventional' LMA one for the 
reactor antineutrino survival probability as measured by KamLAND.
\end{abstract}

\end{titlepage}   
\renewcommand{\thefootnote}{\arabic{footnote}}
\setcounter{footnote}{0}
\section{Introduction} 

After having determined that the solar neutrino problem is essentially a
particle physics one and neutrinos oscillate \cite{Bahcall:2004ut} -
\cite{Aliani:2004bf}, the next step is to search 
for a possible time dependence of the active solar neutrino flux and to 
investigate its low energy sector ($E<1-2~MeV$) which accounts for more than 
99\% of the total flux. These two issues in association with each other may 
lead to further surprises in neutrino physics, possibly the hint of a sizable 
magnetic moment. Although a lot of effort has been devoted to examining the
possible time modulation of the neutrino flux \cite{Aharmim:2005iu} -
\cite{Caldwell:2003dw}, 
this question remains largely unsettled. The claim made in the
early days \cite{Okun:1986hi, Voloshin:1986ty} of a possible anticorrelation 
of the Homestake event rate \cite{Cleveland:1998nv} with sunspot activity remained 
unproven, as no sufficient evidence was found in its support. More recently the 
Stanford Group has been claiming the existence of two peaks \cite{Sturrock:2000jk} 
in the Gallium data at 55-70 SNU and 105-115 SNU. Moreover, Gallium
experiments \cite{Kirsten:1998py} -
\cite{Gavrin:2005ks}, which have been running since 1990-91 and whose event rates 
are mainly due to $pp$ and $^7Be$ neutrinos (55\% and 25\% respectively), also
show a flux decrease from their start until 2003 \cite{Cattadori}. 
These data are hardly consistent with a constant value and exhibit a discrepancy of 
2.4$\sigma$ between the averages of the 1991-97 and 1998-03 periods (see table I). No 
other experiment sees such variations and none is sensitive to low energy neutrinos with 
the exception of Homestake whose rate contains only 14\% of $^7 Be$. Hence this fact 
opens the possibility that low energy neutrinos may undergo a time modulation partially 
hidden in the Gallium data which may be directly connected in some non obvious way to 
solar activity \cite{Chauhan:2004sf}-\cite{Chauhan:2005ju}. Hence also the prime importance 
of the low energy sector investigation. To this end, in the near future, two experiments, 
Borexino \cite{Arpesella:2001iz} and KamLAND \cite{Araki:2004mb}, will be monitoring 
the $^7 Be$ neutrinos.

\begin{center}
\begin{tabular}{ccc} \\ \hline \hline
Period &  1991-97 (I) & 1998-03 (II) \\ \hline
SAGE+Ga/GNO & $77.8\pm 5.0$ & $63.3\pm 3.6$ \\
Ga/GNO only & $77.5\pm 7.7$ & $62.9\pm 6.0$ \\
SAGE only    & $79.2\pm 8.6$ & $63.9\pm 5.0$  \\ \hline
\end{tabular}
\end{center}

{\it{Table 1 - Average rates for Ga experiments in SNU (see ref.\cite{Cattadori}) 
.}}

\vspace{1cm}
In our former work \cite{Chauhan:2004sf, Chauhan:2005pn} we developed a model 
where active neutrinos are partially converted to light sterile ones in addition to 
LMA conversion at times of strong magnetic field, thus leading to the lower Gallium event  
rate (II). When the field is weaker, LMA oscillations act alone and the higher rate (I) is 
obtained. The resonant conversion from active to sterile neutrinos is originated from the
interaction between the magnetic moment and the solar magnetic field. Its location is 
determined by the order of magnitude of the corresponding mass squared difference 
$\Delta m^2_{10}$ and is expected to occur in the tachocline, where a strong time varying 
field is assumed. This implies $\Delta m^2_{10}=O(10^{-8})eV^2$ so that the LMA resonance 
and the active $\rightarrow$ sterile one do not interfere \footnote{For mathematical 
details we refer the reader to \cite{Chauhan:2004sf, Chauhan:2005pn}. We use throughout a
neutrino transition magnetic moment $\mu_{\nu}=10^{-12}\mu_B$.}. It was
shown \cite{Chauhan:2004sf, Chauhan:2005pn} that the high Gallium data as in 
table 1 can be fitted assuming a 'quiet sun' with a weak field while the low data are 
fitted with a strong field ('active sun'). On the other hand the bimodal character of 
the data as claimed by the Stanford Group \cite{Sturrock:2000jk}
cannot be explained in the context of this model. 

The purpose of this paper is to present oscillation and oscillation + spin flavour 
precession (SFP) fits to the data, taking $\Delta m^2_{21},~\theta_{12}$ as free 
parameters. Two separate global fits for Gallium sets (I) and (II) are performed 
with the solar data which were available in the corresponding periods (see table 2). 

\begin{center}
\begin{tabular}{lccc} \\ \hline \hline
Experiment &  Data      &   Theory   &    Reference \\ \hline
Homestake  &  $2.56\pm0.16\pm0.15$ & $8.09\pm^{1.9}_{1.9}$  &
\cite{Cleveland:1998nv} \\
SAGE     &  $see~table~I$ & $125.9\pm ^{12.2}_{12.1}$ &
  \cite{Abdurashitov:2002nt} \\
Gallex+GNO & $see~table~I$ & $125.9\pm ^{12.2}_{12.1}$ &
  \cite{GNO}\\
Kamiokande & $ 2.80 \pm{0.19} \pm{0.33} $ &$5.69\pm{1.41}$&
\cite{Fukuda:1996sz}\\
SuperK&$2.35\pm{0.02}\pm{0.08}$ &
$5.69\pm{1.41}$&
\cite{Fukuda:2002pe}\\
SNO CC &$1.68\pm^{0.06}_{0.06}\pm^{0.08}_{0.09}$&$5.69\pm{1.41}$&
\cite{Aharmim:2005gt} \\
SNO ES &$2.35\pm^{0.22}_{0.22}\pm^{0.15}_{0.15}$&$5.69\pm{1.41}$&
\cite{Aharmim:2005gt} \\
SNO NC & $4.94\pm^{0.21}_{0.21}\pm^{0.38}_{0.34}$&$5.69\pm{1.41}$&
\cite{Aharmim:2005gt} \\ \hline \hline
\end{tabular}
\end{center}
{\it{Table 2 - Data from the solar neutrino experiments except Ga which
is given in Table 1. Units are SNU for Homestake and $10^{6}cm^{-2}s^{-1}$
for Kamiokande, SuperKamiokande and SNO. We use the BS05(OP) solar standard
model \cite{Bahcall:2004pz}.}}

\vspace{1cm}
Hence for data set (I) we consider a rate fit to Gallium, Chlorine and Kamiokande 
data, the only ones existing at the time, together with the KamLAND fit, as these data 
are obviouly independent of solar activity. For set (II) we consider a 
global fit (rates + spectrum) with the exclusion of the Chlorine one, the replacement
Kamiokande $\rightarrow$ SuperKamiokande and the inclusion of the SNO data. We consider
a solar field profile peaked at the bottom of the convective zone. 
We hence determine the parameter values that lead to the best solar+KamLAND 
fit for data set (I). Using these, we establish the best solar fit for set (II) 
on the grounds of its order parameter values $\Delta m^2_{10}$, which situates 
the SFP resonance, and $B_0$, the field strength at the peak. This is also
the best solar+KamLAND fit for data set (II). The Ga (I) and (II) rate 
predictions lie quite close to each other in this fit, both almost 2$\sigma$
away from their central values. This is not totally surprising, as their relative 
weight is small within the wealth of solar+KamLAND data. It tells us instead that
the global fit analysis is not
adequate for the investigation of the possible Ga flux variability. To this end  
we present a choice of a slightly different value of $\Delta m^2_{21},~\theta_{12}$ 
whose fit to data is almost as good as the best fit and
in which the Ga (I) and (II) rate predictions lie much further apart.
We use the BS05 (OP) solar model throughout \cite{Bahcall:2004pz}.

Since our solar fit is independent from the conventional solar one, our only a
priori parameter constraints come from the KamLAND fit for which the mixing angle 
$\theta_{12}$ bears a considerable uncertainty (fig.4 of ref.\cite{Araki:2004mb}). 
In this way we scan the range $\tan^2 \theta_{12}\in[0.1,10]$.
Moreover in the new scenario the prediction for the KamLAND antineutrino 
survival probability is consistent with the measured one \cite{Eguchi:2002dm}. 
A clear distinction between our scenario and the conventional LMA one is expected 
to be provided by the forthcoming KamLAND data if and when new reactors start and 
others cease operation, thus changing the effective source-detector distance travelled 
by the antineutrinos. 

\section{Global Fits and Survival Probability}

We start this section by introducing the following solar field profile
\be
B=\frac{B_0}{ch[6(x-x_{c})]}~~~0<x<x_{c}
\ee
\be
B=\frac{B_0}{ch[15(x-x_{c})]}~~~x_{c}<x<1
\ee
whose peak value $B_0$ is situated at $x_c=0.71$, $x$ denoting the fraction of 
the solar radius. 

For the solar statistical analysis we use the standard $\chi^2$ function
\be
\chi^2_{\odot}=\sum_{j_{1},j_{2}}({R}^{th}_{j_{1}}-{R_{j_{1}}}^{exp})\left[{\sigma^2}
(tot)\right]^{-1}_{j_{1}j_{2}}({R}^{th}_{j_{2}}-{R_{j_{2}}}^{exp})
\ee
where indices $j_{1},j_{2}$ run over all solar neutrino experiments and the error matrix 
includes the cross section, astrophysical and experimental uncertainties \footnote{For 
mathematical details see e.g.\cite{Pulido:1999xp}.}. For the KamLAND 
analysis we compute the prompt energy spectrum of the positron according to 
\be
S(E)=N\int \sigma(E_{\bar\nu})R(E,E^{'})\frac{d\phi}{dE_{\bar\nu}}dE^{'}.
\ee
Here $E^{'}$, $E$ denote the physical and measured prompt event energy, so that $E^{'}$
includes the $e^{+}e^{-}$ annihilation energy and the positron kinetic energy, 
$E^{'}=2m_e+T^{'}_{e^+}$. From the well known relation 
\be
E_{\bar\nu}=m_{N}-m_{P}+m_e+T^{'}_{e^+}=1.804\,MeV+T^{'}_{e^+},
\ee  
(for zero neutron recoil) one gets $E_{\bar\nu}=0.782\,MeV+E^{'}$. Hence eq.(4) is also an 
integral over antineutrino energy. The quantity N is a normalization constant obtained
from the total number of events in the absence of antineutrino disappearance
above the energy cut at 2.6 $MeV$ \cite{Araki:2004mb}
\be
\int S(E) dE=365.2.
\ee
The quantity $\sigma(E_{\bar\nu})$ is the total cross section for the reaction 
$\nu p \rightarrow n e^{+}$ with zero neutron recoil energy \cite{Vogel:1999zy} 
and $R(E,E^{'})$ is the energy resolution function
\be
R(E,E^{'})=\frac{1}{\sigma \sqrt{2\pi}} exp\left[{- \frac{(E-E^{'})^2}{2\sigma ^2}}\right]
\ee
with $\sigma=0.062\sqrt{E}$. The KamLAND energy spectrum comprises 13 bins of size 
0.425 $MeV$ in the range from 2.6 to 8.125 $MeV$. For convenience we integrate eq.(4) 
in antineutrino energy, so that in order to take into account the important low energy 
tail of the spectrum we run the integration from $E_{\bar\nu_{min}}=1.804\,MeV$ up to 
$E_{\bar\nu_{max}}=(8.125+0.782)\,MeV$. In contrast, in eq.(6) the integral extends 
over all 13 energy bins. The information on the flux and spectra of the 20 power 
reactors is contained in the time averaged differential neutrino flux 
$\frac{d\phi}{dE_{\bar\nu}}$ which denotes the number of neutrinos per unit energy, 
area and time. We have used the approximation (see e.g.\cite{Fogli:2005qa}):
\be
\frac{d\phi}{dE_{\bar\nu}}\simeq \sum_{j=1}^{20} P_{osc}(E_{\bar\nu},L_{j})\phi_{j}
\sum_{f=1}^{4}\frac{q_f}{E_f}\frac{dN_f}{dE_{\bar\nu}}.
\ee
In this expression 
\be
P_{osc}(E_{\bar\nu},L_{j})=1-sin^2 2\theta sin^2 \left(\frac{\Delta m^2_{21}L_j}{4E_{\bar\nu}}\right)
\ee
is the well known oscillation survival 
probability formula for antineutrinos from the $j^{th}$ reactor at distance $L_{j}$, 
$\phi_{j}$ is the reactor flux at the KamLAND detector \cite{Gratta}, $q_f$ and $E_f$ 
are the relative fission yields and fission energies \cite{Araki:2004mb, Fogli:2005qa}. 
Following \cite{Fogli:2005qa}, we take
\be
\frac{dN_f}{dE_{\bar\nu}}=exp(a_{f}^{0}+a_{f}^{1}E_{\bar\nu}+a_{f}^{2}E^{2}_{\bar\nu})
\ee
with the coefficients given in \cite{Vogel:1989iv}. 

We have used Poisson statistics for the 13 KamLAND energy bins. Our $\chi^2$ function
therefore is \cite{Bahcall:2004ut}
\be
\chi^2_{KL}=\sum_{i=1}^{13}\left[2(\alpha{R}^{th}_{i}-{R}^{exp}_{i})+2{R}^{exp}_{i}ln
\left(\frac{{R}^{exp}_{i}}{\alpha{R}^{th}_{i}}\right)\right]+\frac{(\alpha-1)^2}{\sigma^2_{sys}}
\ee
with $\alpha$ being an absolute normalization constant and $\sigma_{sys}=6.5\%$ the total
systematic uncertainty. The binned spectrum calculated in the basis of eq.(4) 
added to the background spectrum \cite{KamLAND} is thus inserted in (11) as $R^{th}_{i}$.

We are now in a position to obtain the result of the best fits which we show in table 3. 
For sets (I) and (II) we take $f_B=1.0$. We found for set (I)
\be
\Delta m^2_{21}=8.2\times 10^{-5}eV^2,~tan^2 \theta=0.31,~\alpha=1.01 
\ee
We include for comparison the 'KamLAND only' LMA best fit \cite{Araki:2004mb} with all 
Gallium data replaced by their average \cite{Cattadori}
\be
R_{Ga}=68.3\pm 2.9~SNU
\ee
and with parameters \footnote{The 'LMA parameter values' we
consider are the ones fixed from the best fit analysis to KamLAND data only, since
the conventional solar fits are not taken into account in the present work.}
\be
\Delta m^2_{21}=7.9\times 10^{-5}eV^2,~tan^2 \theta=0.46,~B_0=0,~f_B=0.9,~\alpha=1.01.
\ee
\begin{center}
\begin{tabular}{ccccccccccc} \\ \hline \hline
  & Ga & Cl & K (SK) & $\rm{SNO_{NC}}$ & $\rm{SNO_{CC}}$ & $\rm{SNO_{ES}}$ &
$\!\!\chi^2_{rates}\!\!$ & $\chi^2_{{SK}_{sp}}$ & $\chi^2_{{SNO}_{gl}}$ & $\chi^2_{KL}$\\ \hline 
Set (I)  & 71.7 &  2.66 &  2.29 &  &  &  & 3.09 &  &  & 15.3 \\ 
Set (II)  & 69.6 &   &  2.18 & 5.53 & 1.54 & 2.16 & 2.28 & 44.6 & 45.8 
& 15.3 \\ \hline  
LMA  & 64.8 &  2.74 &  2.30 & 5.10 & 1.75 & 2.28 & 0.95 & 45.7 & 43.1 & 14.5 \\ \hline
\end{tabular}
\end{center}
{\it{Table 3 - Best fits to data sets (I), (II) and LMA best fit. For data set (I) only Ga, Cl
and Kamiokande data were available and for set (II) all SuperKamiokande and SNO data were 
available but not Cl, hence the blank spaces. In set (II) only the Ga rate contributes to
$\chi^2_{rates}$. Units are SNU for Ga and Cl and $10^{6}cm^{-2}s^{-1}$ for SK and SNO.}}

\vspace{1cm}
Using $\Delta m^2_{21},~tan^2 \theta$ from (12), we get the best fit for set (II) 
with
\be
\Delta m^2_{10}=-6.5\times10^{-8}eV^2,~300kG.
\ee
Table 3 $\chi^2_{{SK}_{sp}}$ contains the contribution from electron scattering in 
SuperKamiokande (44 data points) and $\chi^2_{{SNO}_{gl}}$ contains the whole SNO data set consisting 
of 38 data points (34 CC, 2 NC and 2 ES day-night rates). We have 
\be
\chi^2_{gl}=\chi^2_{\odot}+\chi^2_{KL}
\ee
with
\be
\chi^2_{\odot}=\chi^2_{rates}+\chi^2_{{SK}_{sp}}+\chi^2_{{SNO}_{gl}}.
\ee
For sets (I), (II) and LMA we have therefore $\chi^2_{gl}/d.o.f.=1.42,~1,15~1.12$ for
13, 94 and 93 degrees of freedom (d.o.f.) respectively. We thus see that for the 
best global fit the parameter $\Delta m^2_{10}$ is well above $O[10^{-8}eV^2]$. 
This implies that the resonances of the low energy neutrinos lie too deep inside
the sun for the possible time variation of the field to significantly modulate their 
flux. This is particularly true for the $pp$ sector.
In fact the resonances of these neutrinos lie around $x=0.45$, where the field is 
40\% of its maximum and the matter density is much higher than near the peak. Hence 
for the best global fit only a small difference is expected between the two Gallium 
values (see table 3). 

It should be noted however that the global best fit presented here refers to 97 different
experiments and the Gallium data account for only one of them, so their relevance is
smeared by all other data. For a global solar fit the situation is much the same, as the 
number of experiments is reduced to 84 or 16 with the information on the low energy neutrinos 
included only in one of them.  The global fit is therefore not a good measure for the 
investigation of time variability of low energy neutrinos: the information on them is 
nearly hidden within the wealth of solar and KamLAND data.

A small change in the parameters provides a considerable difference: in table 4 
we present the fits for 
$\Delta m^2_{10}=-1.7\times10^{-8}eV^2$, $f_B=1.0$, $f_{Be}=1.1$, $\alpha=0.99$
and $B_0=280kG$ for set (II). This leads to a much larger separation between high and 
low Ga rates in better accordance with table 1 at the price of a slightly 
higher $\chi^2_{gl}$. From table 4 we obtain 
$\chi^2_{gl}/d.o.f.=1.39,~1.19$ for sets (I) and (II) with 13 and 94 d.o.f. respectively.
All values are well within $2\sigma$ of the experimental data. 
\begin{center}
\begin{tabular}{ccccccccccc} \\ \hline \hline
  & Ga & Cl & K (SK) & $\rm{SNO_{NC}}$ & $\rm{SNO_{CC}}$ & $\rm{SNO_{ES}}$ &
$\!\!\chi^2_{rates}\!\!$ & $\chi^2_{{SK}_{sp}}$ & $\chi^2_{SNO}$ & $\chi^2_{KL}$\\ \hline
Set (I)  & {\bf 73.8} &  2.71 &  2.29 &  &  &  & 2.80 &  &  & 15.3 \\
Set (II)  & {\bf 60.3} &   &  2.28 & 5.65 & 1.59 & 2.25 & 0.54 & 47.5 & 48.5
& 15.3 \\ \hline
\end{tabular}
\end{center}
{\it{Table 4 - Same as table 3 with $\Delta m^2_{10}=-1.7\times 10^{-8}eV^2$,
$~B_0=280kG$.}}

\vspace{1cm}
In fig.1  we plot the 'KamLAND only' fit on which we superimpose our own fit
(see table 3) in the plane $\Delta m^2_{21},~tan^2 \theta$. The contour lines correspond
to the 95\%, 99\% , 99.73\% CL. We also show our choice of parameters (table 4) in 
this figure whose goodness of fit is seen to lie well within the 95\% CL. On the other hand a 
comparison with fig.4 (a) of ref. \cite{Araki:2004mb} shows
that it lies inside the 95\% CL of the conventional solar fit
which we neglected in the present paper.
The best fit for the 'low' Ga rate in the plane $\Delta m^2_{10},~B_0$ together with
the 90\%, 95\% and 99\% CL contours is plotted in fig.2 where we also show the
parameter choice of table 4 lying well within the 90\% CL.


We next describe a crucial consistency test for the present scenario, namely the survival 
probability prediction for reactor antineutrinos as measured by KamLAND
\cite{Araki:2004mb}. Our results are depicted in fig.3 for the LMA+SFP fit 
prediction \footnote{We remind the reader that the KamLAND survival probability does
not depend on $\Delta m_{10}$ nor $B_0$.}
with $\Delta m^2_{21}=8.2\times 10^{-5}eV^2,~tan^2 \theta=0.31$ (upper curve) 
and for the LMA best fit, $\Delta m^2_{21}=7.9\times 10^{-5}eV^2,~tan^2 \theta=0.46$
(lower curve). For the average source-detector distance of 180 km, as reported by KamLAND, 
we get from fig.3,
\be
P=0.576~({\rm LMA}),~0.623~({\rm SFP~fit})
\ee
to be compared with the data, $P=0.658\pm0.064$ \cite{Araki:2004mb}.
Interestingly enough it is seen that the best of the two fits is the one for SFP with 
the parameter choice given in table 4, namely the one leading to the two separate Gallium 
data sets at 73.8 and 60.3 SNU. The LMA solution provides the poorest
fit, at 1.28$\sigma$ away from the central value. The present data may only
allow us to expect a clear distinction between the fit of table 4 and the LMA one for
average distances below 110-120 km (see fig.3), which will happen only if new reactors 
come into operation or others cease.
Altogether the quality of the LMA and LMA+SFP energy spectra is
similar ($\chi^2_{KL}/d.o.f.=1.12$ and $1.18$ respectively),  
so that the two predictions are equivalent. 
However the accumulation of more data from KamLAND 
will no doubt clarify the situation. Besides this, we also need more data from the 
low energy solar sector to tell us whether its time modulation is a true or just an
apparent effect.

\section{Summary and Concluding Remarks}

We have considered an alternative to the conventional solar neutrino data fit which
takes into account the possible time dependence of the Gallium flux in the form of
two data sets corresponding to two different periods and separated by 2.4$\sigma$ 
(table 1). This is interpreted as an indication of a possible time variability in the
low energy neutrinos, mainly $pp$, which, though constituting more than 99\% of the 
solar flux, only contribute with 55\% of the Gallium flux.

The simplest mechanism that explains the discrepancy of the two data sets is based on the
partial conversion of active to sterile neutrinos through the spin flavour precession
originating from a varying magnetic field in addition to the LMA oscillation. In the 
global best fit that takes into account all solar data including KamLAND, the two Gallium
rate predictions for each period lie much closer to each other than indicated by the data.
Low energy neutrinos however contribute to only part of the
Gallium rate which in turn amounts to one single experiment within the wealth of solar+
KamLAND data. It is therefore quite natural to expect that the minimum $\chi^2_{gl}$ is 
not obtained for a Gallium rate prediction close to its central value. For this reason 
the global fit analysis is not a good measure for the investigation of the time 
variability of the low energy neutrinos. We have presented a slightly different 
parameter choice with the same $\Delta m^2_{21}$ and $\theta$ values
as the best LMA+SFP fit
(see fig.1) but with  $\Delta m^2_{10}=-1.7\times 10^{-8}eV^2,B_0=280kG$ (see fig.2)
and for which the two Gallium rates are much separated from each other (see table 4). 
It can be seen from figs.1 and 2 that this parameter choice lies near the edge
of the 90\% CL of the best fit. 

The KamLAND spectrum prediction associated with the choice of parameters of table 4 
is practically equal in quality to the LMA one as seen from a comparison between 
tables 3 and 4.
We also tested the antineutrino survival probability as a function of the source-detector
distance. It was found that the parameter choice leading to the two clearly separate
Gallium rates (table 4) provides the survival probability prediction which is in best 
agreement with the data (fig.3).

A close inspection of fig.3 tells us that with the presently existing data, one can only
hope for a distinction between the two scenarios (LMA and LMA+SFP) if the effective
source-detector distance is reduced to less than 110 - 120 km. The Shika2 reactor, which
started operation in late 2005 and gradually increased the operation time at the 
nominal power (4GW thermal power) from last November, will reduce this effective distance 
from 160 - 190 km to 140 - 170 km \cite{Shirai}. Although this cannot provide a definitive 
answer, it is expected that the situation will continue to evolve as new reactors come 
into operation and others cease. The accumulation of more data from KamLAND and 
especially on the low energy solar neutrino sector is essential to disentangle the
prevailing mysteries of solar neutrinos.

\vspace{1cm}
{\Large \bf Acknowledgements}
\vspace{0.5cm}

{\em 
We are grateful to Junpei Shirai for useful discussions and for providing us with
valuable information on the KamLAND experiment. We also acknowledge a discussion with
Mariam Tortola. The works of BCC and MP were supported by Funda\c{c}\~{a}o para a 
Ci\^{e}ncia e a Tecnologia through the grants SFRH/BPD/5719/2001 and SFRH/BPD/25019/2005.
}


\vspace{2cm}

\begin{figure}[h]
\setlength{\unitlength}{1cm}
\begin{center}
\hspace*{-1.6cm}
\epsfig{file=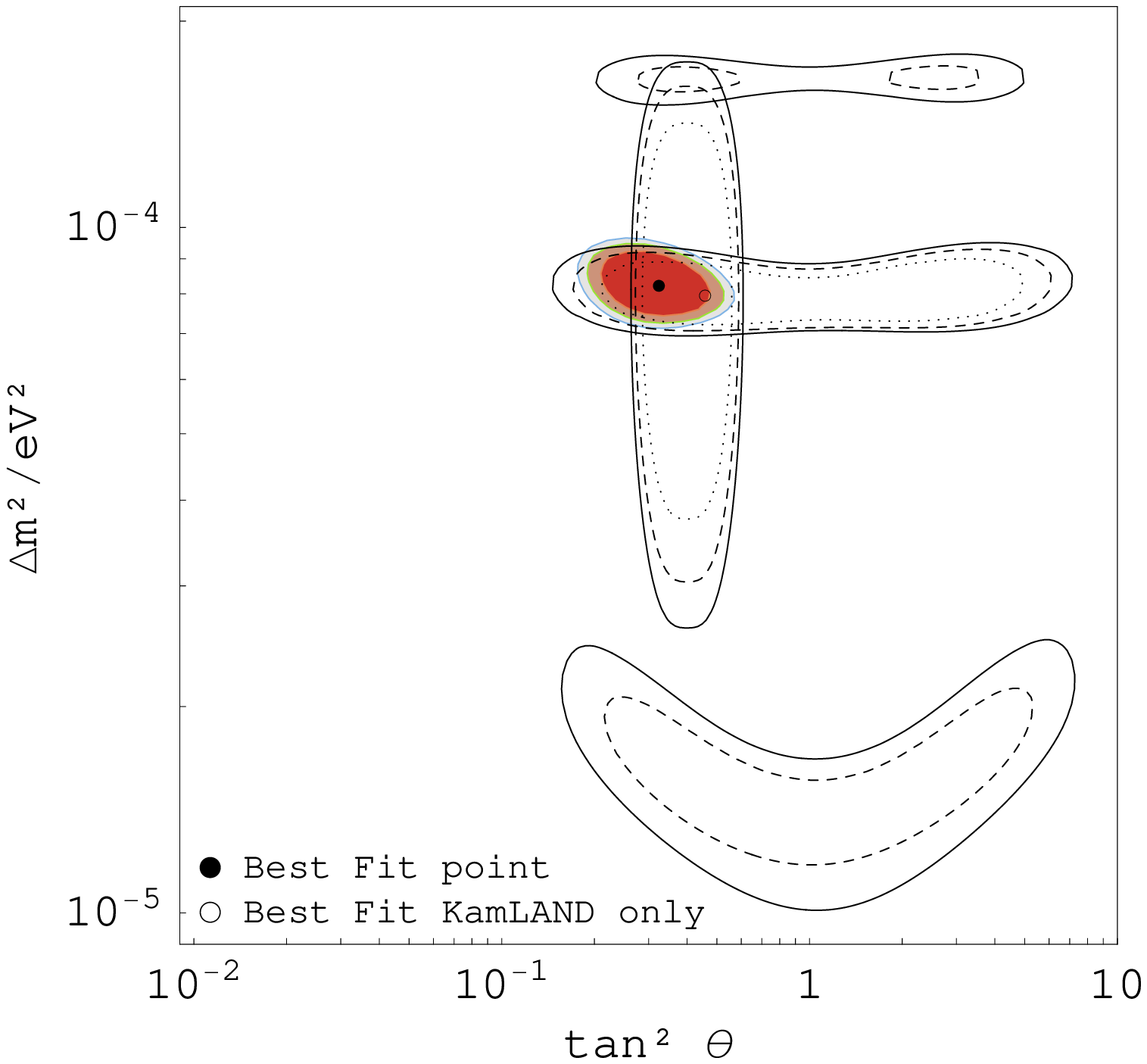,height=14.0cm,width=15.0cm,angle=0}
\end{center}
\caption{ \it Our best fit $\Delta m^2_{21}=8.2\times 10^{-5}eV^2,~tan^2 
\theta=0.31$
(table 3) is superimposed on the neutrino oscillation parameter allowed region from KamLAND 
\cite{Araki:2004mb}. The contour lines are the 95\%, 99\%, 99.73\% CL.
}
\label{fig1}
\end{figure}

\begin{figure}[h]
\setlength{\unitlength}{1cm}
\begin{center}
\hspace*{-1.6cm}
\epsfig{file=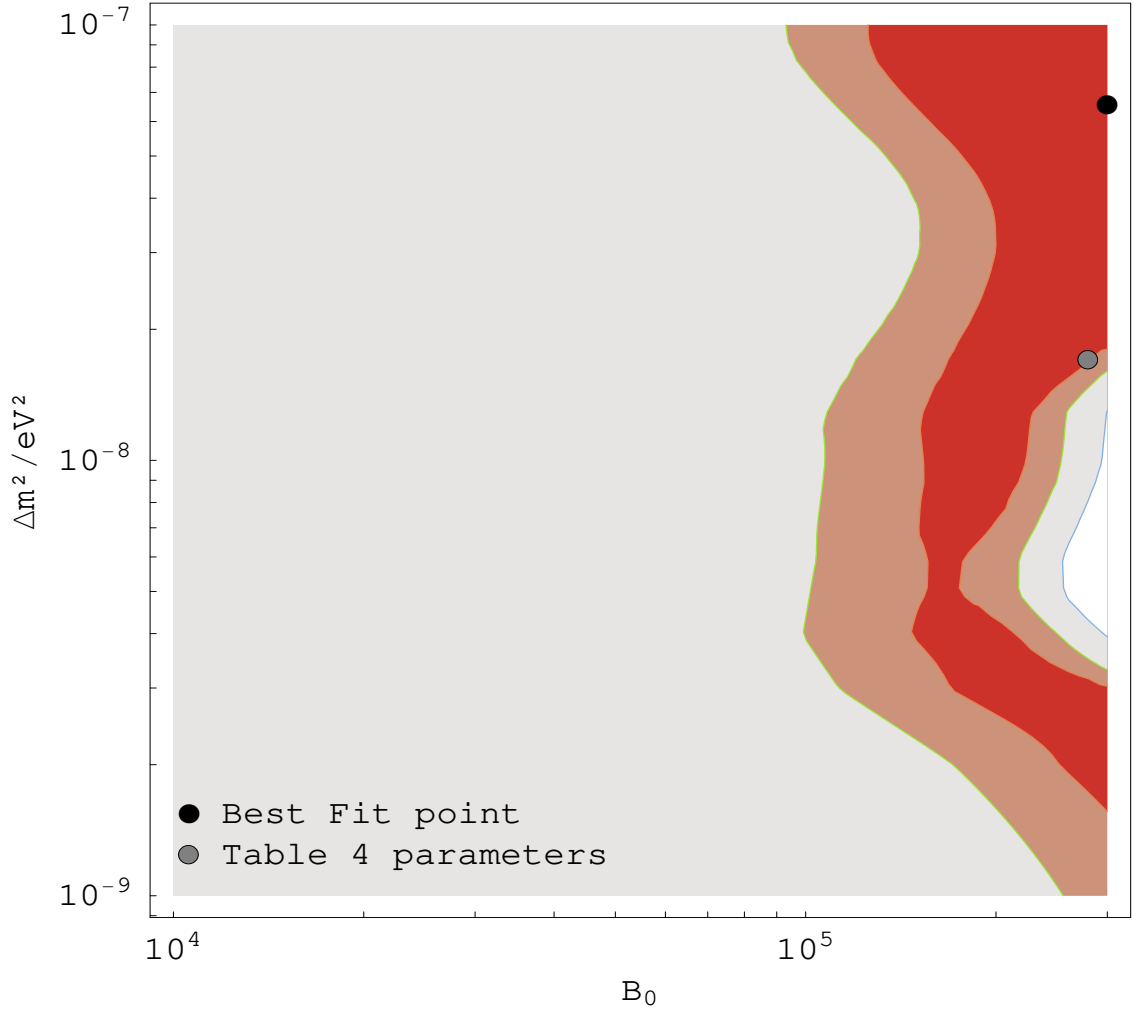,height=14.0cm,width=15.0cm,angle=0}
\end{center}
\caption{ \it The best fit (table 3) in the plane $\Delta m^2_{10},~B_0$ and the  
fit of table 4. The contour lines are the 90\%, 95\%, 99\% CL.}
\label{fig2}
\end{figure}


\begin{figure}[h]
\setlength{\unitlength}{1cm}
\begin{center}
\hspace*{-1.8cm}
\epsfig{file=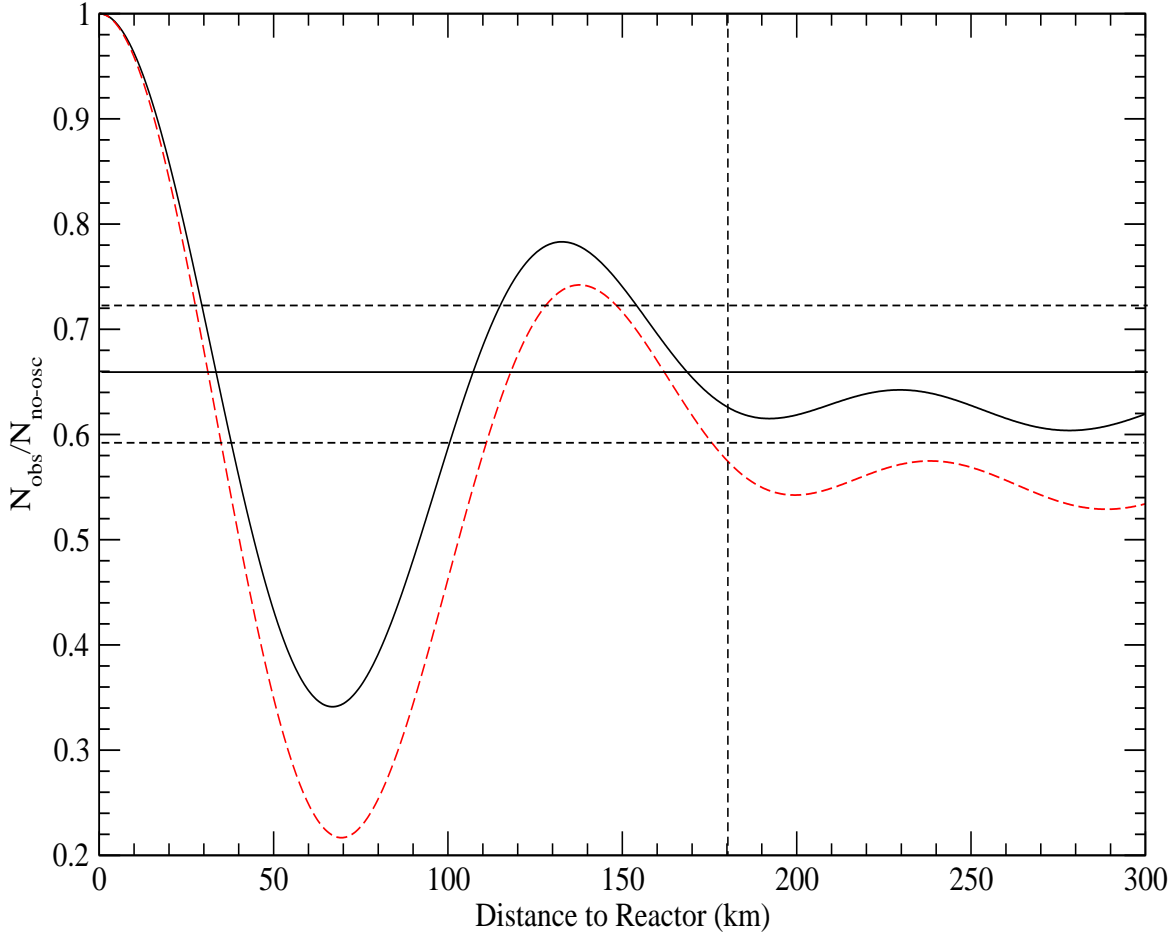,height=18.0cm,width=16.0cm,angle=270}
\end{center}
\caption{ \it The antineutrino survival probability for the fits considered:
LMA (dashed line, table 3) and LMA+SFP ones (full line, tables 3 and 4). Note that 
the two LMA+SFP fits only differ from each other through the values of $\Delta m^2_{10}$ 
and $B_0$ from which the survival probability is independent. Here the straight solid 
line refers to the central data point (P=0.658) and the two horizontal dashed lines to 
the 1$\sigma$ range \cite{Eguchi:2002dm}. The effective reactor distance (180km) is
marked by the vertical dashed line.}
\label{fig3}
\end{figure}

\end{document}